\documentclass[conference]{IEEEtran}
\IEEEoverridecommandlockouts
\usepackage{cite}
\usepackage{amsmath,amssymb,amsfonts}
\usepackage{graphicx}
\usepackage{textcomp}
\usepackage{xcolor}
\usepackage{orcidlink}
\usepackage{url}
\usepackage{stfloats}
\usepackage{booktabs}
\usepackage{listings}%
\usepackage{verbatim}
\usepackage{filecontents}
\usepackage{tabularx}
\usepackage{subcaption}
\usepackage{caption}
\usepackage{tikz}
\usepackage{algorithm} 
\usepackage{algpseudocode} 

\usepackage{listings}   

\usepackage[absolute,overlay]{textpos}
\def\BibTeX{{\rm B\kern-.05em{\sc i\kern-.025em b}\kern-.08em
    T\kern-.1667em\lower.7ex\hbox{E}\kern-.125emX}}
\begin{document}



\title{AI-assisted JSON Schema Creation and Mapping\\
\thanks{Deutsche Forschungsgemeinschaft (DFG) under project numbers 528693298 (preECO), 358283783 (SFB1333), and 390740016 (EXC2075)}
}

\author{\IEEEauthorblockN{1\textsuperscript{st} Felix Neubauer \hypersetup{pdfborder={0 0 0}}\orcidlink{0009-0008-5367-2034}, 3\textsuperscript{rd} Benjamin Uekermann \hypersetup{pdfborder={0 0 0}}\orcidlink{0000-0002-1314-9969}}
\IEEEauthorblockA{\textit{Institute for Parallel and Distributed Systems} \\
\textit{University of Stuttgart}\\
Stuttgart, Germany \\
Felix.Neubauer@ipvs.uni-stuttgart.de,\\
Benjamin.Uekermann@ipvs.uni-stuttgart.de}
\and
\IEEEauthorblockN{2\textsuperscript{nd} Jürgen Pleiss \hypersetup{pdfborder={0 0 0}}\orcidlink{0000-0003-1045-8202}}
\IEEEauthorblockA{\textit{Institute of Biochemistry and Technical Biochemistry} \\
\textit{University of Stuttgart}\\
Stuttgart, Germany \\
Juergen.Pleiss@itb.uni-stuttgart.de}
}

\maketitle

\begin{abstract}
Model-Driven Engineering (MDE) places models at the core of system and data engineering processes.  
In the context of research data, these models are typically expressed as schemas that define the structure and semantics of datasets. 
However, many domains still lack standardized models, and creating them remains a significant barrier, especially for non-experts.
We present a hybrid approach that combines large language models (LLMs) with deterministic techniques to enable JSON Schema creation, modification, and schema mapping based on natural language inputs by the user.
These capabilities are integrated into the open-source tool \textit{MetaConfigurator}, which already provides visual model editing, validation, code generation, and form generation from models.
For data integration, we generate schema mappings from heterogeneous JSON, CSV, XML, and YAML data using LLMs, while ensuring scalability and reliability through deterministic execution of generated mapping rules.
The applicability of our work is demonstrated in an application example in the field of chemistry.
By combining natural language interaction with deterministic safeguards, this work significantly lowers the barrier to structured data modeling and data integration for non-experts.
\end{abstract}


\section{Introduction}

Model-Driven Engineering (MDE) emphasizes the central role of models in the design, development, and operation of systems~\cite{schmidt2006guest}.
In the context of research data management~\cite{pryor2012managing,Wall02012017,hartl2021nationale}, models are typically expressed as schemas, which define the structure and semantics of datasets.  
These models enable key functionalities such as instance validation, transformation, code generation, and visualization, and are essential for ensuring data quality, interoperability, and reusability.


Despite their importance, many scientific domains still lack standardized data models. 
In fields such as chemistry, data are often stored in electronic lab notebooks (ELNs) or spreadsheets, where much of the meaning is implicit rather than explicitly structured.
While CSV documents are used in many machine learning workflows, they are inherently limited to flat, table-like structures and cannot encode relationships, constraints, or domain-specific rules.
This lack of structure restricts interoperability, reproducibility, and the application of downstream methods such as knowledge graph construction or automated schema-driven validation.

Recent advances in large language models (LLMs) offer a promising opportunity to bridge this gap by assisting users, particularly non-experts, in the creation and refinement of data models.
LLMs can translate natural language descriptions into structured model representations, potentially democratizing model-driven techniques. 
However, relying solely on LLMs introduces several limitations: 1) lack of guaranteed model validity, 2) limited interpretability of plain-text outputs, 3) challenges in processing large or complex datasets, and 4) the need for users to craft effective prompts.

To address these issues, we present an MDE-based approach that integrates LLMs with deterministic safeguards, including targeted pre‑ and post‑processing, and the rule‑based execution of AI‑generated mappings, implemented in the open‑source tool \textit{MetaConfigurator}\footnote{\url{https://github.com/MetaConfigurator/meta-configurator}}~\cite{metaconfigurator}. MetaConfigurator supports interactive visual model (schema) editing, validation, and code/documentation/form generation.

We extend the tool with capabilities for schema creation, modification, and querying from natural language input (Section~\ref{sec:standardization}).
By integrating prompt engineering, context management, and post-processing safeguards with visual feedback and schema validation, we align LLM-generated content with the principles of model correctness and transparency central to MDE.
Second, we introduce a model-driven approach to data integration (Section~\ref{sec:integration}): given heterogeneous data sources in JSON~\cite{crockford2006application}, YAML~\cite{ben2009yaml}, XML~\cite{yergeau2006extensible}, or CSV, LLMs generate human-readable mapping rules to transform the data into a target schema (model).
These rules are executed deterministically, separating generation from execution and ensuring reliability and scalability, especially for large datasets.
To demonstrate practical applicability, Section~\ref{sec:application} showcases an example from the domain of chemistry, where existing unstructured Excel data are transformed into structured, interoperable, and AI-ready representations using the extended toolchain.

\section{Background and Related Work}\label{sec:research} 

This section introduces JSON, JSON Schema and work related to schema creation, schema mapping, large language models (LLMs) and prompt engineering.

\subsection{JSON and JSON Schema}
JSON is a widespread data format for exchanging information with web services, storing semi-structured documents in NoSQL databases~\cite{marrs2017json}, and representing structured content in APIs and configuration files.
Its simplicity and human-readability, combined with wide support across programming languages, make it a natural fit for data interchange and machine learning pipelines.

JSON Schema, the de-facto standard for describing the structure and constraints of JSON documents~\cite{baazizi2021empirical}, plays a crucial role in enabling validation, interoperability, and automation.
It can also be understood as a modeling language, capturing the conceptual structure of data much like class diagrams in MDE.

\subsection{JSON Schema Creation}
There exist several so-called JSON schema editors, which are tools for creating and editing schemas.
Among others, this includes MetaConfigurator~\cite{metaconfigurator}, Adamant~\cite{ChaeronySiffa2022} and Liquid Studio JSON Schema Editor\footnote{\url{https://www.liquid-technologies.com/json-schema-editor}, acc. 25/06/02} (paid).
All these editors require some level of understanding of JSON schema.
For this work, we build on MetaConfigurator a general-purpose schema editor and form generator that supports different data formats (e.g., JSON, YAML, XML) and different ways to present and edit the data (e.g., a text editor, GUI editor or schema diagram~\cite{metaconfigurator2025datamodels}).
We choose MetaConfigurator, because of its modular architecture and because it can generate source code in 17 programming languages, as well as generate documentation from a schema.
Furthermore, it is open source, free, accessible as a web service and we are familiar with it.

There also exist approaches to generate a schema automatically, based on an instance dataset.
They are referred to as \textit{schema inference}~\cite{baazizi2019parametric} or \textit{schema discovery}~\cite{spoth2021reducing} tasks. 
LLM-based services, such as ChatGPT, can also directly be used to generate a schema using natural language, however these services lack proper schema validation and visualization of the schema in a graphical way. Furthermore, 1) LLMs can be distracted by irrelevant contexts~\cite{shi2023large}; 2) Even for deterministic tasks, LLMs showed a significant drop in accuracy when dealing with low-probability inputs~\cite{mccoy2023embers}; and 3) The reasoning abilities of LLMs degrades as the input length increases, also before reaching maximum context window~\cite{levy2024same}.
Mior~\cite{mior2024large} fine-tune a LLM for schema related tasks and outperform the base model Code LLama~\cite{roziere2023code} significantly.



\subsection{Schema Matching and Mapping}

Integrating data from various sources and in different formats poses a significant challenge~\cite{putrama2024heterogeneous, ahamed2016data, rozony2024systematic}.
In this work, we study the task of schema mapping: converting an instance from one JSON schema to another.
A schema mapping can contain simple property-to-property correspondences or also more expressive logics.
JSON to JSON transformation (Jolt)\footnote{\url{https://github.com/bazaarvoice/jolt}, acc. 25/06/02}, is a Java library to transform JSON documents. 
JSONata\footnote{\url{https://github.com/jsonata-js/jsonata}, acc. 25/06/02}, is another transformation library, written in TypeScript.
The JSON query language jq\footnote{\url{https://github.com/jqlang/jq}, acc. 25/06/02} also can be used to transform JSON documents.
While it is a powerful tool for concise JSON querying and transformation on the command line, it is not primarily intended for transforming large JSON documents due to its in-memory processing model and limited support for streaming or parallel execution.

Schema mappings can be created manually, but there also exist automated approaches~\cite{hai2018nested}.
The task of identifying which elements in one schema correspond to elements in another is called schema matching.
Rahm and Bernstein~\cite{rahm2001survey} compare different automated schema matching approaches for relational databases.
Stanek and Killough~\cite{stanek2024synthesizing} created a program which generates code to convert a JSON document from one schema to another, defining and implementing different mapping rules themselves. 
Buss et al.~\cite{buss2025scalableschemamappingusing} discuss the use of LLM to automatically create schema mappings, focusing on relational databases.
Among other points, they state that due to the difficult nature of data integration, human-in-the-loop approaches are required.

\subsection{Large Language Models (LLMs) and Prompt Engineering}

LLMs are powerful deep learning models trained on large corpora of text to perform a wide range of natural language processing tasks~\cite{brown2020language,achiam2023gpt,touvron2023llama}.
Prompt engineering is the task of optimizing the input prompt for an LLM to get the best results from it.
Sahoo et al.~\cite{sahoo2024systematic} performed a systematic survey of prompt engineering techniques.
For example, for new tasks without extensive prior training, few-shot training (providing the model a few input-output examples) improves model performance on complex tasks, but requires additional tokens and can lead to biases.
Assigning the LLM a role or persona can also increase the model performance~\cite{kong2023better}.
For complex long instructional prompts, reframing them into multiple smaller subtasks has shown to improve the results~\cite{mishra2021reframing}.
Prior work has shown that providing detailed natural-language instructions significantly improves task performance, compared to relying on implicit task formulation~\cite{ouyang2022training,wang2022self,wei2021finetuned}.

\section{Design and Implementation}\label{sec:design}
In this section, we describe the design and implementation of our AI-assisted schema creation (Section~\ref{sec:standardization}) and schema matching (Section~\ref{sec:integration}) approaches.

Our system communicates with LLMs via a configurable endpoint that follows the OpenAI API\footnote{\url{https://platform.openai.com/docs}, acc. 25/06/07}.
Users may select the desired model via the application's settings.
Prompts are programmatically constructed and transmitted to the API, and the resulting textual outputs are automatically parsed and processed by our application.
Access to LLMs via these APIs requires an authentication key tied to a user account and billing configuration.
We do not provide such API keys directly; instead, MetaConfigurator requires users to supply their own credentials for API access.
All LLM-based interactions and evaluations presented in this paper were conducted using \texttt{gpt-4o-mini}, which we selected for its favorable trade-off between cost-efficiency and its demonstrated proficiency in handling JSON Schema tasks.

\subsection{AI-assisted JSON Schema Creation}\label{sec:standardization}

MetaConfigurator's modular and extensible architecture enables the integration of a new \textit{AI assistance} view.
To enable schema creation and modifications via natural language, we introduce a conversational, chat-like interface.
Our hybrid method offers several advantages over pure LLM-based solutions:
\begin{enumerate}
	\item \textbf{Prompt construction and context management}: The tool handles all aspects of prompt engineering. For schema creation, it dynamically constructs a structured prompt that sets the LLM's role (e.g., \textit{"You are a JSON Schema expert"}), includes the user's natural language description, and specifies the expected output format. For schema modifications, the relevant schema subset, based on the user's current selection, is included in the prompt.
	\item \textbf{Integrated validation and visualization}: The generated or modified schema is immediately subject to validation and is visualized within MetaConfigurator's schema editor.

	\item \textbf{Scalability through targeted context}: To prevent LLM inaccuracies and hallucinations on large schemas, we avoid transmitting the entire schema. Instead, only the user-selected sub-schema is provided as context, enabling precise and modular editing.

	\item \textbf{Automated response post-processing}: The tool automatically cleans the LLM response by removing formatting artifacts such as code fences or language identifiers.

	\item \textbf{Human-in-the-loop editing}: If the LLM output is incomplete or incorrect, the user is presented with the raw response. They may then correct the schema manually before accepting or discarding the change.
\end{enumerate}

\begin{figure}[t]
  \centering

  \begin{minipage}[t]{0.95\linewidth}
    \centering
    \vbox{
      \begin{lstlisting}[breaklines=true,
                         breakatwhitespace=true,
                         linewidth=\linewidth,
                         basicstyle=\ttfamily\scriptsize,
                         breakindent=0pt]
"Create a schema about MOF synthesis in chemistry. We have a list of synthesis experiments. Each experiment has a metal salt and a ligand and also creator, date, temperature, duration, product_purity (boolean). Ligand and metal salt should be objects which each have name, mass, inchi as properties."
      \end{lstlisting}
    }
  \end{minipage}

  \vspace{-0.5em}
  \begin{minipage}{0.05\linewidth}
    \centering
    \Huge$\Downarrow$
  \end{minipage}
  \vspace{0.2em}

  \begin{minipage}[t]{0.95\linewidth}
    \centering
    \includegraphics[width=\linewidth]{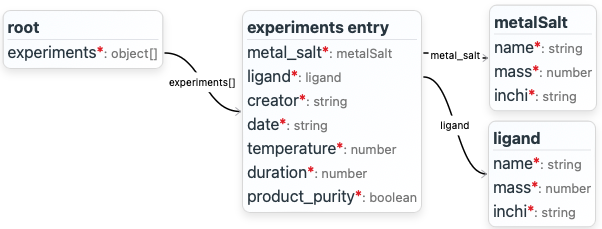}
  \end{minipage}

  \caption{Example of natural language schema creation: The user provides a prompt (top), which is translated into a structured visual schema (bottom).}
  \label{fig:schema_creation_example}
\end{figure}

\begin{figure}[t]
  \centering

  \begin{minipage}[t]{1\linewidth}
    \centering
    \vbox{
      \begin{lstlisting}[breaklines=true,
                         breakatwhitespace=true,
                         linewidth=\linewidth,
                         basicstyle=\ttfamily\scriptsize,
                         breakindent=0pt]
"The metal salt and ligand both have the same structure, both are compounds. Instead of having two different definitions, create one compound class with their properties and then make metal_salt and ligand both refer to this compound class."
      \end{lstlisting}
    }
  \end{minipage}

  \vspace{-0.5em}
  \begin{minipage}{0.05\linewidth}
    \centering
    \Huge$\Downarrow$
  \end{minipage}
  \vspace{0.2em}

  \begin{minipage}[t]{1\linewidth}
    \centering
    \includegraphics[width=\linewidth]{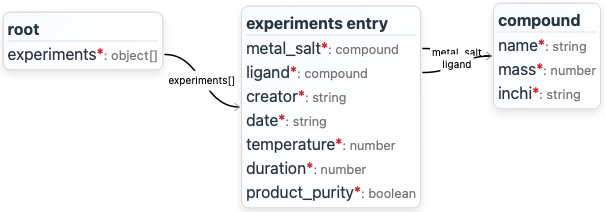}
  \end{minipage}

  \caption{Example of schema modification through natural language: The user requests schema changes using a prompt (top), which results in the updated schema (bottom).}
  \label{fig:schema_edit_example}
\end{figure}

For schema creation tasks, prompts are constructed according to the principles described above.
Figure~\ref{fig:schema_creation_example} shows an example of a user-provided schema description to MetaConfigurator and the resulting schema, visualized.
The same chat-based interface can be used to modify existing schemas (Figure~\ref{fig:schema_edit_example}), to query a schema for information or to create, edit and query document instances.

\subsection{AI-assisted Schema Matching and Mapping}\label{sec:integration}
Schema mapping is performed using human-defined mapping rules or scripts, traditional automated approaches~\cite{hai2018nested, stanek2024synthesizing} and recently also using LLMs~\cite{buss2025scalableschemamappingusing}. 
We propose and implement a new approach, which transforms a JSON document to satisfy a given target schema by first generating schema mapping rules with the help of a LLM and then executing these deterministically. 
Aside from JSON, MetaConfigurator supports the data formats YAML and XML and has a function to import CSV documents (by conversion to JSON).
Therefore, data from any of these formats are supported and can be transformed using the approach.

As schema mapping language, we use JSONata.
It is very expressive and supports complex features, such as numeric, path, boolean and comparison operators, sorting, grouping, aggregation, functions and expressions, functional programming and higher order functions.

To perform the schema matching and mapping (data model transformation), the user provides an input JSON document and a target JSON schema.
First, a source schema is inferred for the provided JSON document instance, using the jsonhero/schema-infer\footnote{\url{https://github.com/triggerdotdev/schema-infer}, acc. 25/06/05.} library.
Next, the prompt for the LLM is constructed.
It consists of mapping instructions (giving the LLM a personality, describing the overall tasks and detailed mapping rules), a mapping example (example input and expected output files) and the user input.
To improve the quality of the generated JSONata expressions, a detailed set of instructions and JSONata specifications\footnote{Our set of JSONata instructions: \url{https://doi.org/10.18419/DARUS-5157}} is included.

Because the user input JSON document might be large, we perform a recursive array and properties truncation to shorten the document while maintaining its overall structure.
The algorithm iteratively trims all arrays and object properties to a maximum length $n$ and $8*n$ respectively, until either the target document size of \texttt{64KB} or the minimum $n_{min}=2$ is reached.
The value \texttt{n} starts at \texttt{64} and every iteration it is divided by $2$.
Properties of objects are trimmed more conservatively than array items, to a length of $8*n$, as they are usually all relevant and different (except for schemas with \texttt{additionalProperties} or \texttt{patternProperties}), contrary to arrays items, which tend to follow the same schema.
Possible information loss is constrained by including the JSON schema inferred from the complete JSON document in the prompt.
Because of including the whole document schema in the prompt, there is an upper boundary of document complexity which is supported.

\begin{figure}[t]
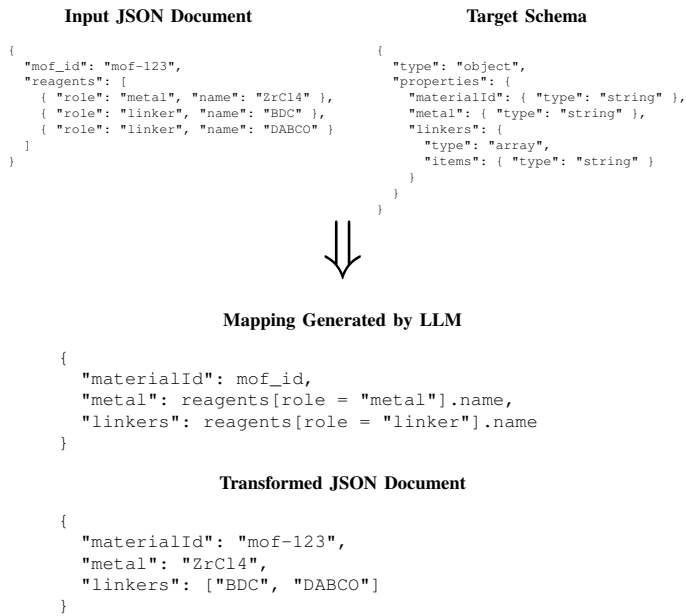

  \centering

  \begin{minipage}[t]{0.45\linewidth}
    \centering
    \scriptsize
    \textbf{Input JSON Document}
    
    \begin{lstlisting}[firstnumber=1, linewidth=\linewidth,
                       basicstyle=\ttfamily\tiny]
{
  "mof_id": "mof-123",
  "reagents": [
    { "role": "metal", "name": "ZrCl4" },
    { "role": "linker", "name": "BDC" },
    { "role": "linker", "name": "DABCO" }
  ]
}
    \end{lstlisting}
  \end{minipage}
  \hfill
  \begin{minipage}[t]{0.45\linewidth}
    \centering
    \scriptsize
    \textbf{Target Schema}
    
    \begin{lstlisting}[firstnumber=1, linewidth=\linewidth,
                       basicstyle=\ttfamily\tiny]
{
  "type": "object",
  "properties": {
    "materialId": { "type": "string" },
    "metal": { "type": "string" },
    "linkers": {
      "type": "array",
      "items": { "type": "string" }
    }
  }
}
    \end{lstlisting}
  \end{minipage}

  \vspace{-0.5em}
  \begin{minipage}{\linewidth}
    \centering
    \Huge$\Downarrow$
  \end{minipage}
  \vspace{0.2em}

  \begin{minipage}[t]{0.85\linewidth}
    \centering
    \scriptsize
    \textbf{Mapping Generated by LLM}
    
    \begin{lstlisting}[firstnumber=1, linewidth=\linewidth,
                       basicstyle=\ttfamily\scriptsize]
{
  "materialId": mof_id,
  "metal": reagents[role = "metal"].name,
  "linkers": reagents[role = "linker"].name
}
    \end{lstlisting}

    \vspace{0.2em}
    \scriptsize
    \textbf{Transformed JSON Document}
    
    \begin{lstlisting}[firstnumber=1, linewidth=\linewidth,
                       basicstyle=\ttfamily\scriptsize]
{
  "materialId": "mof-123",
  "metal": "ZrCl4",
  "linkers": ["BDC", "DABCO"]
}
    \end{lstlisting}
  \end{minipage}

  \caption{Example of schema mapping: given an input JSON document and a target schema (top), the LLM generates a transformation expression (middle), which is then used to derive the structured output (bottom).}
  \label{fig:schema_mapping_example}
\end{figure}

After the prompt is constructed, sent to the LLM and a response is received, a post-processing algorithm is executed on the response.
The response might be surrounded by a code fence, often with a language hint (\texttt{jsonata}).
If applicable, it is removed.
Finally, the resulting suggested schema mapping is presented to the user in an interactive code editor.
It is automatically validated for syntactical correctness (not semantics) using the JSONata library\footnote{\url{https://github.com/jsonata-js/jsonata}, acc. 25/06/09.} and validation errors are shown to the user.
The user can make any changes of their choice and can only apply the mapping once it is recognized as valid.
Figure \ref{fig:schema_mapping_example} shows an example schema mapping input, the generated JSONata expression and the transformation result.

\section{Application Example}\label{sec:application}

Chemists have performed chemical synthesis experiments and tracked their inputs and results in an Excel table.
They want to apply machine learning and other algorithms on their data, but it is not yet in a suitable format\footnote{Sources: \url{https://doi.org/10.18419/DARUS-5157}.}.
MetaConfigurator is used to convert the excel table into a JSON document and automatically infer the corresponding schema.
The JSON document and schema, however, are not yet of sufficient quality.
Among other points, the \texttt{product\_purity} values of \texttt{yes} and \texttt{no} have been inferred as being of type \texttt{string} instead of \texttt{boolean} and the \texttt{ligand} and \texttt{metal\_salt} properties are all listed directly as children of an experiment instance, although it would be more suitable to extract them into separate \texttt{ligand} and \texttt{metal\_salt} classes.
Using the chat-based interface we create an improved schema (Figure~\ref{fig:schema_creation_example}-\ref{fig:schema_edit_example}).
To transform the JSON document to satisfy the new schema, we use the automated schema mapping functionality.
It successfully maps from \texttt{yes} and \texttt{no} to boolean values and from the flat source paths to the nested paths of the target schema.
The data is now well-structured, formally specified by a schema and AI-ready.

For the automated execution of experiments using laboratory robots, existing standards are preferred.
For example, the creators of Chemspyd~\cite{chemspyd}, a python interface for operating laboratory robotic platforms from Chemspeed Technologies, provide the outlook of integrating the Chemical Description Language XDL~\cite{mehr2020universal} into Chemspyd.
As of writing this paper, no formal schema is available for XDL 2.0, aside from documentation\footnote{\url{https://croningroup.gitlab.io/chemputer/xdl/standard}, acc. 25/07/05.}.
We copy and paste the relevant sections of the XDL documentation into the schema creation prompt of MetaConfigurator to generate a schema.
Since XDL documents fundamentally differ from our data in structure, JSONata cannot provide an adequate mapping.
Therefore, we use the code generation capabilities of MetaConfigurator to generate python classes representing the data structure of our schema and the XDL schema and write our own python code to convert between the structures.
The result is a XDL document instance that is compatible with the XDL ecosystem and might enable an automatic synthesis execution in the near future.

\section{Conclusion}

We presented a hybrid approach for schema creation, editing, and instance transformation using large language models, integrated into the open-source tool MetaConfigurator. 
Our method combines modular prompt engineering, targeted context scoping, human-in-the-loop refinement, and post-processing safeguards to enable intuitive and controlled schema modeling.
To enable transforming also large document instances to satisfy a target schema, mapping rules are generated by LLMs but then applied in a deterministic manner.

An example in the chemistry domain demonstrates the applicability of our approach.
By translating informal data into machine-readable structures, our method enables FAIR (Findable, Accessible, Interoperable, and Reusable)~\cite{wilkinson2016fair} data practices and model-driven workflows. 

Beyond this proof-of-concept, model editing and instance transformation, the tool supports other MDE-oriented tasks such as instance validation, code/documentation generation, mapping discovery and schema-driven form generation.
A future extension for model-to-model transformations (e.g., XSD to JSON Schema) is planned.

\section*{Acknowledgment}

The authors acknowledge Esengül Ciftci (Max Planck Institute for Solid State Research, Stuttgart, Germany) and Kenichi Endo (University of Stuttgart, Institute of Polymer Chemistry, Stuttgart, Germany) for inspiring the chemistry application example.
The assistance of ChatGPT-4 for editorial suggestions is acknowledged.

\end{document}